# Performance limitation of Si Nanowire solar cells: Effects of nanowire length and surface defects

Deepika Bora[1,3], Shrestha Bhattacharya[2,3], Nitin Kumar[3,4], Aishik Basu Mallick[2,3], Avriti Srivastava[3,4], Mrinal Dutta[3]*, Sanjay K. Srivastava[3,4], P. Prathap[3,4] and C.M.S. Rauthan[3]

[1]Amity Institute of Applied Science, Amity University, Noida, India
[2] Department of Energy Engineering, Central University of Jharkhand, Jharkhand-835205, India
[3]PV Metrology Group, Advanced Materials Devices and Metrology Division, CSIR-National Physical Laboratory (NPL), New Delhi-110012, India
[4] Academy of Scientific and Innovative Research, CSIR-NPL Campus, New Delhi-110012, India

**Corresponding author: sspmd.iacs@gmail.com, duttamrinal1081@gmail.com**

**Abstract:** In Si nanowire (SiNW) solar cells enhanced light confinement property in addition to decoupling of charge carrier collection and light absorption directions plays a significant role to resolve the draw backs of bulk Si solar cells. In this report we have studied the dependence of the phovoltaic properties of Si NW array solar cells on the SiNW length and enhanced surface defect states as a result of enhanced surface area of the NWs. The SiNW arrays have been fabricated using metal catalyzed electroless etching (MCEE) technique. p-n junction has been produced by spin-on-dopant technique followed by thermal diffusion process. Front and rear electrodes have been deposited by e-beam evaporation techniques. SiNW lengths have been controlled from ~ 320 nm to 6.4 μm by controlling the parameters of MCEE technique. Photovoltaic properties of the solar cells have been characterized by measuring quantum efficiency and photocurrent density vs. voltage characteristics. Morphological studies have been carried out by using scanning electron microscopy. Reduction in light trapping capability comes at the benefit of reduced surface defects. The reduction of surface defects has been proved to be more advantageous in comparison to the decrement of light trapping capability. The major contribution to the changes in cell efficiency comes from the enhancement of short circuit current density with a very weak dependence on open circuit voltage. This work is beneficial for the production commercial Si solar cell where SiNW arrays could be used as a antireflection coating instead of using separate antireflection layers and thus could reduced the production cost.

## Introduction:

Silicon nanowires (SiNWs) have attracted global attention as a promising material to achieve high efficiency at low cost due to their unique structural, optical, and electrical properties [1, 2]. As Si has low absorption in the visible and near infrared region of the solar spectrum, so efficient commercial Si solar cells to fully absorb incident sunlight need relatively large amounts of high-purity solar-grade Si. In this decade, worldwide effort has been started to address this problems by fabricating SiNW based solar cells. Due to light-trapping within the NW arrays these cells exhibit a higher absorbance per unit thickness compare to commercial Si solar cells and thus open a path to avoid the need of extra antireflection coating layer that could lower the production cost [3].

There are several processes like CVD, molecular beam epitaxy, reactive ion etching in addition to lithography, metal catalyzed electroless etching (MCEE) technique etc. which are used to fabricate SiNWs [4]. Among these MCEE technique is only simple and low-cost while others are costly and time consuming. By MCEE technique SiNWs could be fabricated in wafer scale and the electrical characteristics of the SiNWs are same as mother Si wafer. Formation of homo or hetero junction could be achieved either by thermal diffusion or deposition of opposite polarity shell layer over the NWs in core-shell morphology by CVD technique or by spin coating opposite polarity polymer layer. Thermal diffusion involved high temperature processing while in CVD technique low temperature processing could be used. On the other hand polymer coated heterojunction cells most of the time are prepared at room temperature but subjected to reliability and stability issues. Low crystalline quality of the shell layers deposited by CVD technique most of the time degraded the performance of SiNW solar cells produced in core-shell morphology. In these SiNW solar cells surface states plays a dominant role. With increase of surface area surface related defect recombination starts to take over other types of recombination process (e.g. bulk recombination). Reduction of surface roughness helps in reducing surface recombination. This has direct effect on cell efficiency by improving efficiency from 0.5% to 5.3% [5, 6]. On the other hand insertion of thin intrinsic-Si layer improves the efficiency up to 7.29% [7]. Reduction of length of the SiNWs also showed improvement of cell efficiency up to a certain height of the NWs. In this study p-n junctions were fabricated in core-shell morphology by depositing p-Si shell layer over n-SiNWs by cold walled thermal CVD [6]. The control of surface states of the SiNW array solar cells fabricated by thermal diffusion process has rarely been investigated. The simple way to control these surface states is to shorten the NW length that has been proved in our previous work [6].

Here we report a control over the surface states by controlling the NW height for the thermally diffused SiNW solar cells. Without using any surface or interface passivation layer this investigation shows more than 30% enhancement in cell efficeincy. Thus through this work we have established a compromise between enhanced junction area and increased surface defect states with increase of NW length. This study is worth for the commercial Si solar cell production where SiNW arrays could be used as an light trapping layer instead of depositing any extra antireflection coating layers.

## Experimental:

SiNW arrays were prepared by MCEE technique on 300 μm thick p-type (100) single side polished crystalline p-Si substrates with resistivity of 5-10 ohm-cm. The Si substrates were cleaned with acetone (10 min) and ethanol (10 min) by sonication and then rinsed with deionized water 2–3 times and boiled in a 3:1 mixture of $H_2SO_4$ (96%) and $H_2O_2$ (30%) for 15 min and rinse by (deionised) D. I. Water thoroughly. After dipping in 2% HF for 2 minute immediately transferred to a Teflon beaker containing a solution of 0.04M of $AgNO_3$ in 7.8% of HF for specific time to form vertically aligned Si NW arrays. The substrates were then rinsed thoroughly with DI water. The etched Si substrates were immersed in a solution of ammonia and hydrogen peroxide in 7:3 ratio for 10 min to remove Ag den-drites which had been deposited during etching. Again the water treatment is done 2-3 times. After this nitric acid treatment was done in order to remove Ag remnant (if any).

The P507 (phosphorous dopant solution) has been spin coated on the SiNW arrays and planar substrate. Before spinning the phosphorous containing solution on the NWs, cleaning of SiNW arrays were done by using acetone for 5 min, then by ethanol for 5 min. After this substrates are boiled in piranha solution for 15 min at 70˚C. The substrates were then washed with water for several times then dipped in 2% HF and throughly rinsed with de-ionized water twice. After the spin-coating the substrates were heated at 90˚C for 10 min on the hot plate and at $200^0C$ in oven for 15 min. After that we used the diffusion furnace to diffuse P on the front surface at 990˚C for a chosen time period. Diffusion of dopant creates PSG layer. Now to remove PSG these Si substrates were dipped in 2% HF solution for 5 min and thoroughly washed by deionized water.

For fabrication of Al electrode on the back surface electron beam evaporator at a base pressure~ $3\times10^{-6}$ mbar was used. Thickness of the electrode was kept around 1 μm. After deposition of back contact BSF formation was achieved. Ag electrode in finger grid pattern was deposited by electron beam evaporator at a base pressure~ $3\times10^{-6}$ mbar. Contact sintering was done after that to achieve ohmic contact. The effective area of the solar cells was kept at 1 $cm^2$.

## Characterization

A Scanning Electron Microscope (SEM, Model: ZEIS EVO MA 10) was used for microstructural characterization. J-V and EQE characterizations were performed through M/s Bunkoukeiki system.

## Results and Discussion:

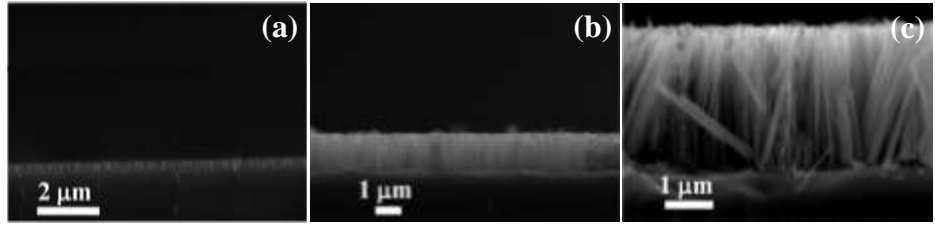

Figure 1. Cross sectional SEM images of the SiNW array solar cells having SiNW length of (a) 320 nm, (b) 1550 nm, (c) 6400 nm

Figure 1 shows the cross-sectional images of the SiNW arrays solar cells having different NW lengths from 320 nm to 6400 nm. External quantum efficiency (EQE) of the SiNW array based cells and planar cell are shown in Fig. 2. A spectrally broad EQE has been observed for the planar cell. The EQE shows increment for the all SiNW array cells compare to planar cell with some special features. It has been found that at short wavelengths EQE significantly decreased with increasing NW length. With increase of NW length roughness increases as NWs surface are rough. This increase of roughness with increase of surface area increases the surface related dangling bond defects that enunciate more surface recombination with increase of NW length [6]. In our previous study reported elsewhere it was shown that with increase of SiNW length light trapping increases and as a result reflection reduces more and more with increase of SiNW length [6]. However for longer wavelengths EQE of SiNW array solar cells show some enhancement that could be ascribed to enhance light trapping. With increase of NW length at longer wavelength due to light trapping EQE increases up to a certain amount then it decreases with more increase in NW length. This has happened because of the fact that with increase of NW length surface related recombination starts to overcome the advantage of excellent light trapping property of longer wavelengths. Figure 3 shows the photo-current density vs voltage characteristics of SiNW array solar cells and planar solar cell under AM1.5G solar illumination (100 mW/cm$^2$). The solar cell parameters of these cells are presented in Table 1. From Table 1 it could be observed that SiNW array improves the cell performance compare to planar cell but with increase of NW length the cell performance degraded. For NW length up-to 1550 nm cell performance is still higher than that of planar cell. However more increase in NW length degraded the cell performance and it goes below even that of planar cell for NW length 6400 nm.

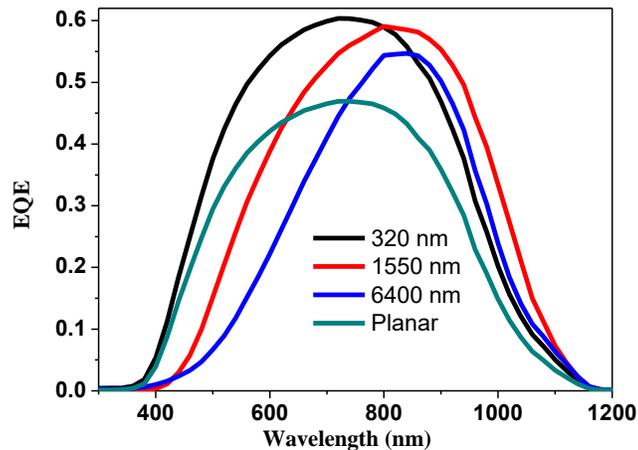

Figure 2. EQE of the SiNW array solar cells and planar solar cell

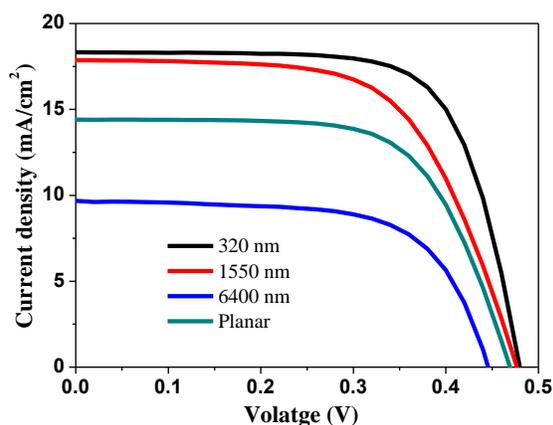

Figure 3 Photocurrent density vs. voltage characteristics of planar cell and SiNW solar cells having different NW lengths under simulated AM1.5G illumination

Table 1. Solar cell parameters of SiNW array and planar cell

| NW length (nm) | $J_{SC}$ (mA/cm$^2$) | $V_{OC}$ (V) | FF | η (%) |
|---|---|---|---|---|
| 0(planar) | 14.41 | 0.47 | 0.65 | 4.45 |
| 320 | 18.32 | 0.48 | 0.7 | 6.16 |
| 1550 | 17.82 | 0.477 | 0.62 | 5.27 |
| 6400 | 9.67 | 0.445 | 0.65 | 2.98 |

## Conclusion:

SiNW arrays of different NW length were fabricated by simple and low-cost MCEE technique. p-n junction was fabricated by spin-on-dopant technique followed by thermal diffusion of dopants. Current-voltage and EQE measurements shows that for SiNW arrays with smaller length of the NWs cell performance is higher than the planar cells while for longer NW length the cell performance degraded and even go below that for planar cell.

## Acknowledgement:

Authors want to thank DST-SERB Ramanujan Fellowship programme for providing financial support and also CSIR-NPL for providing instrument facility.